\documentstyle[aas2pp4,adjworkshop]{article}

\begin{document}

%
%
%
   \def\prd#1#2#3#4{#4 19#3 Phys.~Rev.~D,\/ #1, #2 }
   \def\pl#1#2#3#4{#4 19#3 Phys.~Lett.,\/ #1, #2 }
   \def\prl#1#2#3#4{#4 19#3 Phys.~Rev.~Lett.,\/ #1, #2 }
   \def\pr#1#2#3#4{#4 19#3 Phys.~Rev.,\/ #1, #2 }
   \def\prep#1#2#3#4{#4 19#3 Phys.~Rep.,\/ #1, #2 }
   \def\pfl#1#2#3#4{#4 19#3 Phys.~Fluids,\/ #1, #2 }
   \def\pps#1#2#3#4{#4 19#3 Proc.~Phys.~Soc.,\/ #1, #2 }
   \def\nucl#1#2#3#4{#4 19#3 Nucl.~Phys.,\/ #1, #2 }
   \def\mpl#1#2#3#4{#4 19#3 Mod.~Phys.~Lett.,\/ #1, #2 }
   \def\apj#1#2#3#4{#4 19#3 Ap.~J.,\/ #1, #2 }
   \def\aj#1#2#3#4{#4 19#3 Astr.~J.,\/ #1, #2}
   \def\acta#1#2#3#4{#4 19#3 Acta ~Astr.,\/ #1, #2}
   \def\rev#1#2#3#4{#4 19#3 Rev.~Mod.~Phys.,\/ #1, #2 }
   \def\nuovo#1#2#3#4{#4 19#3 Nuovo~Cimento~C,\/ #1, #2 }
   \def\jetp#1#2#3#4{#4 19#3 Sov.~Phys.~JETP,\/ #1, #2 }
   \def\sovast#1#2#3#4{#4 19#3 Sov.~Ast.~AJ,\/ #1, #2 }
   \def\pasj#1#2#3#4{#4 19#3 Pub.~Ast.~Soc.~Japan,\/ #1, #2 }
   \def\pasp#1#2#3#4{#4 19#3 Pub.~Ast.~Soc.~Pacific,\/ #1, #2 }
   \def\annphy#1#2#3#4{#4 19#3 Ann. Phys. (NY), \/ #1, #2 }
   \def\yad#1#2#3#4{#4 19#3 Yad. Fiz.,\/ #1, #2 }
   \def\sjnp#1#2#3#4{#4 19#3 Sov. J. Nucl. Phys.,\/ #1, #2 }
   \def\astap#1#2#3#4{#4 19#3 Ast. Ap.,\/ #1, #2 }
   \def\anrevaa#1#2#3#4{#4 19#3 Ann. Rev. Astr. Ap.,\/ #1, #2
                       }
   \def\mnras#1#2#3#4{#4 19#3 M.N.R.A.S.,\/ #1, #2 }
   \def\jdphysics#1#2#3#4{#4 19#3 J. de Physique,\/ #1,#2 }
   \def\jqsrt#1#2#3#4{#4 19#3 J. Quant. Spec. Rad. Transfer,\/ #1,#2 }
   \def\jetpl#1#2#3#4{#4 19#3 J.E.T.P. Lett.,\/ #1,#2 }
   \def\apjl#1#2#3#4{#4 19#3 Ap. J. (Letters).,\/ #1,#2 }
   \def\apjs#1#2#3#4{#4 19#3 Ap. J. (Supp.).,\/ #1,#2 }
   \def\apl#1#2#3#4{#4 19#3 Ap. Lett.,\/ #1,#2 }
   \def\astss#1#2#3#4{#4 19#3 Ap. Sp. Sci.,\/ #1,#2 }
   \def\nature#1#2#3#4{#4 19#3 Nature,\/ #1,#2 }
   \def\spscirev#1#2#3#4{#4 19#3 Sp. Sci. Rev.,\/ #1,#2 }
   \def\advspres#1#2#3#4{#4 19#3 Adv. Sp. Res.,\/ #1,#2 }
   %
%
%
\def\Msun{M_{\odot}}
\def\Mdot{\dot M}
\def\deg{$^\circ$\ }
\def\etal{{\it et~al.\ }}
\def\eg{{\it e.g.,\ }}
\def\etc{{\it etc.}}
\def\ie{{\it i.e.,}\ }
\def\ksec{{km~s$^{-1}$}}
\def\arcsec{{$^{\prime\prime}$}}
\def\arcmin{{$^{\prime}$}}
\def\subsun{_{\twelvesy\odot}}
\def\sun{\twelvesy\odot}
\def\gtwid{\mathrel{\raise.3ex\hbox{$>$\kern-.75em\lower1ex\hbox{$\sim$}}}}
\def\ltwid{\mathrel{\raise.3ex\hbox{$<$\kern-.75em\lower1ex\hbox{$\sim$}}}}
\def\plusminus{\mathrel{\raise.3ex\hbox{$+$\kern-.75em\lower1ex\hbox{$-$}}}}
\def\minusplus{\mathrel{\raise.3ex\hbox{$-$\kern-.75em\lower1ex\hbox{$+$}}}}

\title{Lyman Edge Features from Accretion Disks Around Maximally Rotating
Supermassive Black Holes}

\author{Mark W. Sincell}
\affil{DAEC, Observatoire de Meudon\\
92195 Meudon cedex, France\\ 
{\tt sincell@mesioc.obspm.fr}}

\begin{abstract} 
We calculate the amplitude of the Lyman discontinuity for an accretion disk
around a maximally rotating black hole.  The locally emitted disk spectrum 
is computed with a previously developed numerical model (Sincell \&
Krolik 1998) and then a new general relativistic ray-tracing 
code is used to determine the observed spectra as a function of inclination
angle, $\theta_o$.  We find that Lyman 
discontinuities are undetectable for disks viewed at inclinations 
$\cos\theta_o \ltwid 0.8$, but a significant feature remains for 
disks seen face-on.

\end{abstract}

\section{Introduction}
\label{sec: introduction}

The ultraviolet (UV) spectrum of Seyfert galaxies is dominated by a
quasi-thermal component (Shields 1978, Malkan \& Sargent 1982, Malkan
1983) which extends into the soft x-ray band (Walter \& Fink 1993).
Although this feature is usually interpreted as thermal emission from a
geometrically thin, optically thick accretion disk around a supermassive
black hole, recent observations cast doubt on this simple explanation.
A disconinuity at the Lyman edge, either in emission or absorption, seems to 
be a generic feature of theoretical models of accretion disks 
(\eg Blaes 1998).  However, partial 
absorption features
have been detected in $\ltwid 10\%$ of observed AGNs (Antonucci, Kinney \&
Ford 1989, Koratkar, Kinney \& Bohlin 1992).  

Recently, Iwasawa, \etal (1996) detected extremely broad iron K-$\alpha$
line emission from the Seyfert galaxy MCG-6-30-15. They interpret this
emission as x-rays reflected by an accretion disk extending to 1.23 $R_g$,
where $R_g = GM/c^2$, of the central black hole.  For a  stable accretion
disk to exist this close to the central black hole this black hole must be
rotating close to its maximal value (\eg
Shapiro \& Teukolsky 1983). 

The appearance of an accretion disk around a maximally rotating Kerr black
hole is very different than that of a disk around a non-rotating hole.
The effective temperature of the disk is larger.  The material in the disk
orbits the hole at higher velocities, increasing the range of Doppler boosts of
the disk radiation.  Finally, the emitted radiation has to escape from a deeper
potential well, which causes larger redshifts.

In this paper, we compute the observed spectrum of a viscously heated
accretion disk around a maximally rotating supermassive black hole.  The
rest frame spectrum of radiation emitted by the disk is computed using the
code developed by Sincell \& Krolik (1998).  The observed disk spectrum is
calculated using a ray-tracing program which we describe in the next section.

\section{The Ray-Tracing Program}

The photon equations of motion in a Kerr metric are derived from the Lagrangean
\begin{equation}
2 L       = g_{tt} \dot x^t \dot x^t + 2 g_{t\phi} \dot x^t \dot x^{\phi}
            + g_{rr} \dot x^r \dot x^r 
            + g_{\theta\theta} \dot x^{\theta} \dot x^{\theta}
            + g_{\phi\phi} \dot x^{\phi} \dot x^{\phi}
\end{equation}
using the geodesic equation
\begin{equation}
\label{eq: lagrangean eq motion}
{d \over d\lambda} \left( {\partial L \over \partial \dot x^i }
\right)
 = {\partial L \over \partial x^i }
\end{equation}
where $\lambda$ is an affine parameter and $i$ indicates the coordinate 
of interest: $ t,r,\theta,\phi$. 
Expressions for the metric components, $g_{ij}$ are given by Bardeen, \etal
(1972) for Boyer-Linquist coordinates. 
Equation \ref{eq: lagrangean eq motion} expands to eight first-order
coupled differential equations for the four space coordinates and the four
components of the photon momentum.

This equation set can be reduced to six first-order equations by noting that
the photon energy
\begin{equation}
E = -p_{t}
\end{equation}
and angular momentum parallel to the symmetry access
\begin{equation}
L = p_{\phi},
\end{equation}
where $p_i = \partial L / \partial \dot x^i$, are
conserved along a null geodesic.  It is possible to reduce the equation set
further (see Bardeen, \etal 1972, Cunningham 1975, Agol 1997) but this 
introduces complicated 
bookkeeping into the ray-tracing calculation. The differential equations
are integrated with a fourth-order Runge-Kutta routine (Press, \etal 1992). 

There is one additional constant of the motion
\begin{equation}
Q^2 = p_{\theta}^2 + cos^2 \theta \left[ - a^2 p_{t}^2 + p_{\phi}^2 
/ sin^2\theta \right]
\end{equation}
which is needed to specify a particular photon geodesic.  In this equation, $a$
is the black hole spin parameter and $\theta =
x^{\theta}$ in Boyer-Lindquist coordinates. 

To compute the observed spectrum, we choose a particular inclination angle
and a grid in the impact parameters
$(\alpha,\beta)$. The parameter $\alpha$, which is measured in units of $R_g$,
is defined as the projected distance away from the 
black hole along an axis perpendicular to the spin axis of the black hole. The
parameter $\beta$ is the projected distance parallel to the spin axis.
The range of values for the impact parameters corresponds roughly to the 
inner and outermost radii of the accretion disk.

For each pair of impact parameters, the constants of motion of the 
corresponding geodesic are defined
\begin{equation}
L = - \alpha \sin\theta_o
\end{equation}
\begin{equation}
Q^2 = \beta^2 + (\alpha^2 - a^2)\cos^2 \theta,
\end{equation}
and $E=1$. The equations of motion are then integrated inward from $r_o \gg
M$, where $M$ is the mass of the black hole and we adopt geometrized units,
 and $\theta=\theta_o$.  The integration is stopped when the geodesic either
hits the disk ($\theta=\pi/2$), escapes to infinity or is captured by the
black hole.

For a geodesic which intersects the disk at radius $r_e$, we can compute two 
important quantities: the total redshift of
a photon emitted by the disk at $r_e$ and escaping to infinity
\begin{equation}
g = {\nu_e \over \nu_o},
\end{equation}
where $\nu_{e,o}$ are the frequency of the emitted and observed photon,
and the angle at which the photon leaves the disk, $\theta_e$, measured in
the rest frame of the orbiting material.  Explicit 
equations for these quantities can be found in Cunningham (1975).

The observed spectrum is found by integrating the emitted intensity over the 
impact parameters of all the geodesics which reach the observer
\begin{equation}
I(\nu_o,\theta_o) = \int I(\nu_e,r_e,\theta_e) g^3 d\alpha d\beta
\end{equation}
where the factor $g^3$ arises from the invariance of the photon
occupation number and $I(\nu_e,r_e,\theta_e)$ is the spectral intensity of 
the radiation emitted by the disk.  
For example, the intensity could be described by an analytic expression 
for the reflected iron line as a function of radius or a grid of numerical 
models of the continuum flux from an optically thick accretion disk.

\section{Results}

\begin{figure}
\plotone{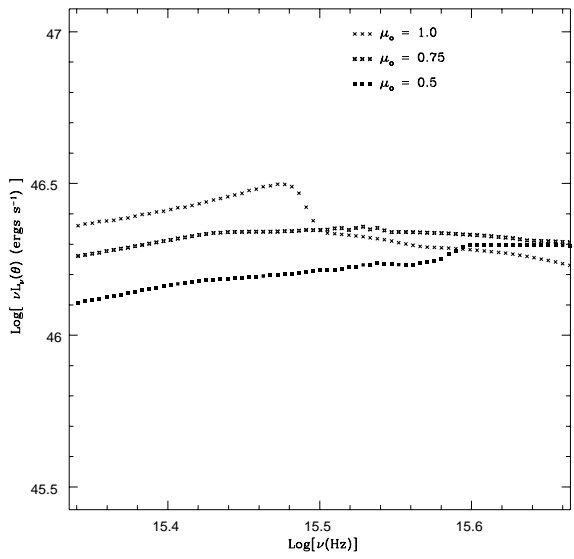}
\caption{\label{fig: continuum} A model of continuum emission from a disk
around a Kerr black hole.}
\end{figure}

In figure \ref{fig: continuum}, we plot the UV continuum of an optically
thick accretion disk around a maximally rotating black hole. The black hole 
is assumed to have a mass of $2.7 \times 10^{8} \Msun$ and to accrete
 matter at $\dot m = 0.3 \dot M_E$, where $M_E$ is the Eddington 
critical accretion rate.   The three curves represent three different values
of the disk inclination angle, $\theta_o = \cos^{-1} \mu_o$.

We find that the Lyman edge discontinuity is unobservable when $\mu_o \ltwid 
0.8$, but that a significant absorption feature remains at smaller 
inclinations.  If this model is typical of all accretion disks, then we would 
predict that Lyman edge features should be seen in $\ltwid 20\%$ of all
AGNs.  This is obviously only a very rough estimate, but it does indicate that
black hole rotation is a plausible means of erasing the Lyman discontinuity
in a large fraction of AGNs.

\end{document}